\documentclass[preprint]{aastex701}
\usepackage{multirow}
\usepackage{placeins}
\usepackage{booktabs}

\newboolean{cleanversion}
\newboolean{redactive} 

\setboolean{cleanversion}{true} 
\setboolean{redactive}{false}    

\newcommand{\rev}[1]{%
  \ifthenelse{\boolean{cleanversion}}%
  {#1}
  {\textcolor{blue}{#1}}
}

\newcommand{\red}[1]{%
  \ifthenelse{\boolean{redactive}}%
  {\textcolor{red}{#1}}
  {#1}
}

\begin{document}

\title{Temporal Evolution of Sunspot Groups and Increase in the Open flux During Solar Maximum in Cycle 24}

\author[orcid=0000-0002-7206-7162]{Minami Yoshida}
\affiliation{Department of Earth and Planetary Science, The University of Tokyo, 7-3-1, Hongo, Bunkyo-ku, Tokyo 113-0033, Japan}
\affiliation{Institute of Space and Astronautical Science, Japan Aerospace Exploration Agency, 3-1-1, Yoshinodai, Chuo-ku, Sagamihara, Kanagawa 252-5210, Japan}
\email{yoshida.minami@ac.jaxa.jp}

\author[orcid=0000-0003-4764-6856]{Toshifumi Shimizu}
\affiliation{Institute of Space and Astronautical Science, Japan Aerospace Exploration Agency, 3-1-1, Yoshinodai, Chuo-ku, Sagamihara, Kanagawa 252-5210, Japan}
\affiliation{Department of Earth and Planetary Science, The University of Tokyo, 7-3-1, Hongo, Bunkyo-ku, Tokyo 113-0033, Japan}
\email{shimizu.toshifumi@jaxa.jp}

\author[orcid=0000-0002-1276-2403]{Shin Toriumi}
\affiliation{Institute of Space and Astronautical Science, Japan Aerospace Exploration Agency, 3-1-1, Yoshinodai, Chuo-ku, Sagamihara, Kanagawa 252-5210, Japan}
\email{toriumi.shin@jaxa.jp}

\author[orcid=0000-0002-1007-181X]{Haruhisa Iijima}
\affiliation{Institute for Space-Earth Environmental Research (ISEE), Nagoya University,  Furo-cho, Chikusa-ku, Nagoya 464-8601, Japan}
\email{haruhisa.iijima@gmail.com}


\begin{abstract}

    \rev{The evolution of the global solar magnetic field directly impacts the interplanetary magnetic field (IMF). During the solar maximum of Cycle 24, the monthly averaged IMF strength doubled over five Carrington rotations in late 2014. To understand the physical origin of this increase, we investigate the temporal evolution of open magnetic flux resulting from the emergence and decay of bipolar magnetic regions (BMRs). Using surface flux transport and potential field source surface models, we simulated how BMR characteristics, spatial distributions, and interaction with background magnetic fields affect open flux evolution. \red{Our simulation confirmed} that the relative configuration of BMRs can either inhibit open flux expansion via closed loops or promote it through favorable connections. The increase in open flux is primarily driven by the equatorial dipole component, which is enhanced by differential rotation acting on tilted BMRs. These behaviors suggest that large open field structures develop from equatorial dipole components formed by these stretched BMRs.
    We attribute the rapid IMF increase in 2014 (Carrington rotations 2152-2157) to the combination of the following three factors: (1) a specific sunspot configuration that facilitated the expansion of the southern coronal hole; (2) the emergence of a giant sunspot group (active region 12192) with high magnetic intensity; and (3) the diffusion of these regions, which reinforced the global magnetic field. These results imply that rapid open flux variations during solar maximum are governed not only by the characteristics of emerging BMRs but also by their interaction with pre-existing large coronal holes.}
     
\end{abstract}

\keywords{Solar magnetic fields (1503); Bipolar sunspot groups(156); Sunspot groups(1651); Solar differential rotation(1996); Interplanetary
magnetic fields (824); Heliosphere (711)}


\section{Introduction}\label{sec:intro}
The magnetic field structure of the Sun changes over the solar cycle. 
During the solar minimum, sunspots rarely appear in the photosphere, and the magnetic field lines extending from the polar regions form a dipole magnetic structure in the corona. These polar regions, called coronal holes, are less bright in soft X-ray and EUV images and are magnetically unipolar \citep{1973SoPh...29..505K}.
During the solar maximum, sunspots appear at the surface at low and middle latitudes, making the coronal magnetic field structure more complex. Changes in the overall coronal structure with the solar cycle have been well captured observationally by coronagraphs  (\citealp{1992SSRv...61..393K}; \citealp{2019ApJ...883..152D}) and during solar eclipses (\citealp{2015ApJ...800...90P}; \citealp{2018NatAs...2..913M}). These changes are also simulated numerically by coronal field modeling using an input of the spatial distribution of magnetic flux observed in the photosphere (\citealp{2014SoPh..289..631Y}, \citealp{Wiegelman&Sakurai2021}).

The magnetic field lines that return to the photosphere are called closed fields, whereas those extending to the interplanetary space are called open fields. The open fields extend dominantly from the coronal holes, according to the standard coronal structure \citep{2012LRSP....9....6M}. These coronal holes have been thought to be the origin of fast solar wind. However, open fields may also exist outside coronal holes. The open field lines are connected to the interplanetary magnetic field (IMF), whose strength also varies with the solar cycle. 
The connection of the solar magnetic field to the IMF is essential for understanding the magnetic structure in the interplanetary space and across the heliosphere. 

The strengths of the IMF and solar wind velocity have been observed in situ near the Earth since 1965 \citep{2013LRSP...10....5O}. Currently, the Wind \citep{1995SSRv...71..207L}, Advanced Composition Explorer (ACE; \citealp{2005GeoRL..3215S01B}), and Deep Space Climate Observatory (DSCOVR) satellites measure them at the Lagrangian point 1 (L1), which are publicly available in the OMNI dataset \citep{https://doi.org/10.1029/2004JA010649}.
The Parker Solar Probe (PSP; \citealp{2016SSRv..204....7F}; \citealp{10.1063/PT.3.5120}), launched in 2018, has a close approach to the Sun approximately every 3 months and has provided in situ measurements at a distance of $10-30$ solar radii from the solar surface. Such satellites provide accurate measurements of the magnetic field and velocity only at the positions where they fly. 

The three-dimensional magnetic field structure in interplanetary space can be inferred only by the extrapolation of solar magnetic fields from the photosphere. 
\citet{2017ApJ...848...70L} pointed out that the magnetic field strength of the open flux extrapolated from the photospheric field is two to five times smaller than the magnetic field strength measured in situ near the Earth. This discrepancy is known as the open flux problem. This problem has not yet been solved, even by applying any different photospheric magnetogram to any kind of coronal modeling  (\citealp{2019ApJ...884...18R}; \citealp{2022ApJ...926..113W}; \citealp{2024ApJ...970..131S}). The discrepancy is also observed at a distance closer to the Sun, as measured by the PSP \citep{2021A&A...650A..19R}. 
Therefore, one of the ultimate reasons we cannot solve the open flux problem is that we have not fully clarified the comprehensive connectivity between the solar surface magnetic field and the IMF.

\begin{figure}[ht]
    \centering
    \includegraphics[width=.9\columnwidth]{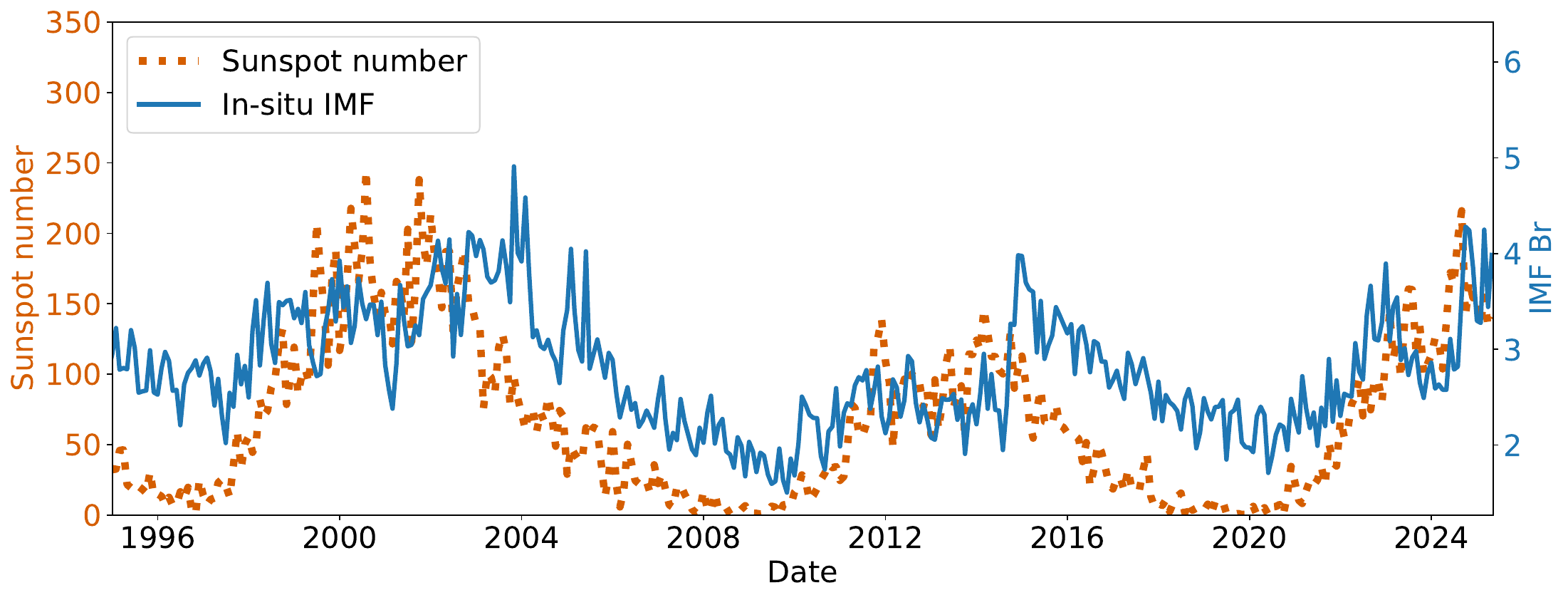}
    \caption{
    Temporal variations in IMF at 1 AU (blue solid line) and sunspot numbers (orange dotted line) during Cycle 23 and Cycle 24 (1-month running average). A characteristic increase in IMF is observed in the latter half of 2014 during Cycle 24. \red{The sunspot number data is provided by WDC-SILSO, Royal Observatory of Belgium, Brussels (DOI: https://doi.org/10.24414/qnza-ac80), and the in-situ IMF data is obtained from the NASA/GSFC's OMNI database.}} 
    \label{fig:imf_sunspot}
\end{figure}

The distribution of open fields during the solar maximum is poorly understood compared to the solar minimum, when coronal holes are concentrated in the polar regions. 
Observational studies have suggested that low-order components of the solar magnetic field are important for increasing open flux and IMF (\citealp{2013ApJ...775..100P}; \citealp{2014SSRv..186..387W}). In particular, the equatorial dipole component is important during the solar maximum (\citealp{2022ApJ...926..113W}; \citealp{2023ApJ...950..156Y}).
\rev{A characteristic IMF temporal variation occurring during the solar maximum is observed in 2014 (Figure \ref{fig:imf_sunspot}.) The approximately twice increase in the monthly-averaged IMF over the 5 Carrington rotations from June to November 2014 (CR 2152 to CR 2156). \citet{2023ApJ...950..156Y} showed this increase in the IMF is consistent with the timing of the peak of the solar equatorial dipole flux when the magnetic field diffuses toward the polar region. \citet{heinemann2024origin} investigated the open flux evolution during part of this period (September to October 2014). They found that the increase in the open flux correlated with the formation of a large southern coronal hole, which they attributed to the poleward flux transport from active regions decaying several months earlier. They also noted a potential link to the emergence of the largest active region in Cycle 24. However, their study remained qualitative regarding the causal link, and it has not been discussed which specific characteristics of sunspot groups, such as their geometric configuration or global magnetic field structures, caused this rapid increase. Therefore, it is necessary to understand how the increase or decrease in total open flux is determined in response to temporal and spatial changes in sunspot groups for one to several Carrington rotations (CRs). To understand this observation, a theoretical understanding of how individual sunspot groups determine open flux is necessary.}

\rev{Theoretical approaches using simplified bipolar magnetic regions (BMRs) provide a solid foundation for understanding the physical mechanism of how sunspot emergence contributes to the increase and decrease of open flux. 
The surface flux transport (SFT) model (e.g., \citealp{1983ApJ...270..288S}; \citealp{1985AuJPh..38..999D}; \citealp{1989ApJ...347..529W}) and potential field source surface (PFSS) model (\citealp{1969SoPh....9..131A}; \citealp{1969SoPh....6..442S}) are widely used to simulate the evolution of the photospheric magnetic flux. 
Pioneering works by \citet{2002JGRA..107.1302W} and \citet{2002SoPh..207..291M} established that the evolution of total open flux is governed by the dipole strength of the photospheric magnetic field. They theoretically demonstrated how specific BMR properties, such as tilt angle (the tilt of the centers of gravity of the leading and following sunspots with respect to the solar latitude) and emergence latitude, and the interaction of dipole vectors from multiple BMRs determine the global field structure.
Specifically, they showed that a larger tilt angle results in a greater latitudinal separation between the leading and following sunspots. Consequently, differential rotation effectively stretches the magnetic flux as it diffuses, leading to an increase in the open flux over time.
They also indicated that BMRs emerging at lower latitudes are more effective in contributing to the global dipole field and maintaining the open flux. 
Furthermore, they pointed out that the spatial distribution of BMRs determines the strength of the total dipole moment, which in turn regulates the fraction of the magnetic field that opens into interplanetary space.} 
Some simulation studies also have investigated in detail how specific BMR parameters control the open flux generation (e.g., \citealp{2006A&A...459..945S}; \citealp{2010ApJ...719..264C}; \citealp{2011A&A...528A..83J}).
\rev{Thus, the application of such forward-modeling techniques has become a well-established framework for deciphering the complex evolution of the global magnetic field observed in reality.}
\rev{However, most of these studies primarily focused on statistical behaviors or long-term variations over the solar cycle. To understand rapid variations such as the 2014 event, it is necessary to clarify how the global open flux is determined in response to the temporal and spatial evolution of sunspot groups over timescales of one to several Carrington rotations (months).}

\rev{In this study, we apply these established theoretical frameworks to the specific rapid variation in the IMF observed in 2014, focusing on the configuration of photospheric magnetic fields (BMRs) and the short-term evolution of the open flux. The evolution of the photospheric magnetic flux is simulated using the SFT model and the PFSS model.
First, the dependence of the open flux evolution on the BMR configuration is quantified by using the SFT model.}
Second, the simulation results are applied to attribute the cause of a rapid increase in the open flux and IMF that was observed in late 2014 during the solar maximum of Cycle 24.
In Section \ref{sec:analysis}, we present the model and analysis. 
In Section \ref{sec:results}, we describe the results and discussion of the time evolution in the open flux using the SFT and PFSS models. \rev{This section is structured into two parts: (1) simulation results for BMRs and surrounding magnetic field case (Section \ref{sec:result2}) and (2) discussion on simulation results (Section \ref{sec:discussion_simu}).} In Section \ref{sec:result_obs}, we focus on the rapid IMF increase in 2014 by presenting the observational results (Section \ref{sec:result:IMFobs}) and applying our simulation results to this event (Section \ref{sec:discussion:IMFobs}.) Finally, in Section \ref{sec:conclusion}, we summarize this study.

\section{Methodology}\label{sec:analysis}
\subsection{Surface Flux Transport Model}
\begin{figure}[ht]
    \centering
    \includegraphics[width=.9\columnwidth]{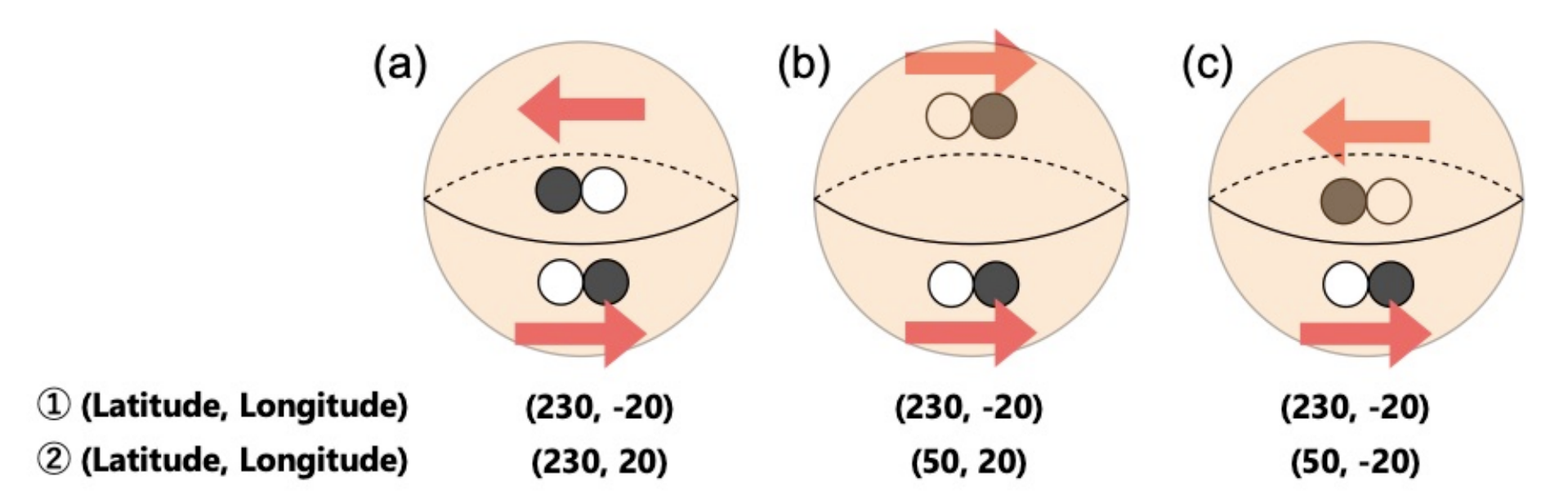}
    \caption{
    Spatial distributions of two BMRs for the three cases of $\#$2. The orange sphere represents the Sun, the white and black circles depict sunspots with positive and negative polarity, respectively, and the red arrow indicates the direction of the magnetic field. (a) Placing two bipolar BMRs at the same longitude. The BMR in the southern hemisphere has a magnetic polarity opposite to that placed in the northern hemisphere. (b) Two bipolar BMRs, one of which is placed at a different longitude (180$^{\circ}$ apart) in a different hemisphere. (c) Two BMRs are placed in the southern hemisphere, one of which is placed apart 180$^{\circ}$ from the other.}
    \label{fig:sun_tilt}
\end{figure}

In this analysis, we used artificial magnetograms with arbitrarily placed BMRs. The temporal evolution of the coronal magnetic field and its open flux were then simulated. The BMR was approximated using the model by \citet{1998ApJ...501..866V}.
\rev{We focused on the spatial configuration and strengths of BMRs as the primary parameters governing the variation of open flux as established in previous studies (Section \ref{sec:intro}). Latitude and tilt angle were fixed to typical values for 2014 (tilt angle was $10^{\circ}$ and latitude was $20^{\circ}$). We defined the maximum magnetic flux density as 2000 G for a ``large BMR" and 1000 G for a ``small BMR," with the size being determined by a Gaussian function and the spatial separation between the leading and following polarities being scaled to this size.
Two categories of magnetogram were created for the SFT simulations. For the first magnetogram, BMRs were placed in a zero initial background magnetic field to validate that our model reproduces fundamental behaviors consistent with previous studies.
For the second magnetogram, new BMRs were placed in a background magnetic field that had already diffused for 3 to 5 months. This setup was created to reproduce the conditions of the large coronal hole observed from August to November 2014. Using this pre-existing background, we performed simulations to evaluate the impact of magnetic flux content by varying BMR intensities (Configuration \#1; a large versus small BMRs) and to test the configuration effect in a realistic environment (Configuration \#2 as shown in Figure \ref{fig:sun_tilt}; simlar configuration to \citealp{2002JGRA..107.1302W}). For these cases, we also analyzed the spatial expansion of the open field footpoints.}

\red{Since the governing transport equation} in the SFT model is linear for the magnetic field, it is possible to examine the direct relationship between the input BMR and the resultant magnetic field evolution and understand how each BMR contributes to the open flux. This linearity helps us to gain insights into the contribution of BMRs to the open flux in the actual, far more complicated Sun.
The number of grids in the magnetogram was set to (longitude, latitude)=(128, 64). 

The two-dimensional SFT code developed by H. Iijima \citep{2019ApJ...883...24I} is used in this study.
\rev{The velocity field consists of the differential rotation $V_{\phi}$ and the meridional flow $V_{\theta}$.
In this study, the differential rotation profile is adopted from \citet{1983ApJ...270..288S}.
The meridional flow is assumed to be steady and uniform in the longitudinal direction.
We used the profile described by \citet{1998ApJ...501..866V} with a peak velocity of approximately $11~\rm{m~s^{-1}}$, following the setup of \citet{2019ApJ...883...24I}.
Turbulent diffusion represents the horizontal diffusion of the magnetic field caused by thermal convection.
While earlier models often used a diffusion coefficient of $600~\rm{km^{2}~s^{-1}}$ (e.g., \citealp{2002JGRA..107.1302W}), we adopted a lower value of $\eta = 250~\rm{km^{2}~s^{-1}}$ which is uniform in both space and time.
This value is consistent with recent SFT simulations (\citealp{2019ApJ...883...24I}) and allows the model to retain smaller-scale magnetic structures.}
The temporal variation was calculated for one year, starting from each of the 74 different photospheric magnetograms as the initial condition. The time resolution of these output magnetograms was 5 days.

\subsection{Potential Field Source Surface Model}
Using a magnetic field map obtained at each step in the time evolution for the 74 cases, we derived the coronal magnetic field by using the PFSS model. 
The PFSS model is one of the simplest coronal magnetic models, based on two major assumptions.
First, it assumes a potential field, namely, that the electric current vanishes in the magnetic field.
The second assumption is that all the magnetic fields are oriented in the radial direction at the source surface, which is the upper boundary. Following previous studies, the source surface was set to 2.5 solar radii (e.g. \citealp{2023ApJ...950..156Y}).
The PFSS code used in this study was developed by M. L. DeRosa (\url{https://www.lmsal.com/~derosa/pfsspack}).

\subsection{Observational Data}

In the discussions on a rapid change observed in the amount of the open flux and IMF in 2014, we used OMNI dataset (\url{https://omniweb.gsfc.nasa.gov/form/dx1.html}) for in-situ measurement data and the Helioseismic and Magnetic Imager (HMI) on board the Solar Dynamics Observatory (SDO) spacecraft (\citealp{2012SoPh..275....3P}; \citealp{2012SoPh..275..207S}) for magnetograms. We obtained synoptic maps created from line-of-sight magnetograms available at the Joint Science Operations Center (JSOC; \url{http://jsoc.stanford.edu}), and reduced the size of the maps to 600 $\times$ 300 pixels. We calculated the total unsigned magnetic flux in the photosphere, the total open flux at the source surface, and the solar equatorial dipole flux. See \citet{2023ApJ...950..156Y} for details of the calculations. IMF is  calculated as $4\pi R^{2}\langle\vert B_{r}\vert\rangle$, where $R$ is the distance from the Sun to the L1 point ($1.496\times 10^{13}$ cm), and $\langle\vert B_{r}\vert\rangle$ is the 27 days average of radial IMF measured at L1. Solar equatorial dipole flux is calculated as a component of $l=1$ and $m=\pm1$ for spherical harmonic function derived from the synoptic map of magnetograms (\citealp{2023ApJ...950..156Y}, Appendix A1). The period examined is from Carrington Rotation (CR) 2152 to 2157, a period during the solar maximum in Cycle 24, during which period we have observed a spiky increase in the IMF. 

\FloatBarrier
\section{Results and Discussion from the simulation}\label{sec:results}
\rev{We show the results of our SFT and PFSS simulations using artificial BMRs. The results of realistic cases with initial background magnetic fields are described in Section \ref{sec:result2}, and we discuss the physical mechanisms governing the results in Section \ref{sec:discussion_simu}.}

\subsection{Result: Effect of BMRs with Initial Background Magnetic Fields on Open Flux}\label{sec:result2}
\rev{We examine the evolution of open flux resulting from the interaction between new BMRs and diffused initial background magnetic fields. The initial magnetograms were created by \red{evolving} a small BMR placed at longitude $230^{\circ}$ and the latitude $-20^{\circ}$ for 3-5 months, and the subsequent evolution is analyzed over 2.5 months to apply the observational timeline.}

\subsubsection{Dependence on the strength of a BMR with Initial Background Magnetic Fields (Configuration $\#$1)}\label{sec:ch_rad}

\begin{figure}[ht]
 \begin{center}
   \includegraphics[width=.45\columnwidth]{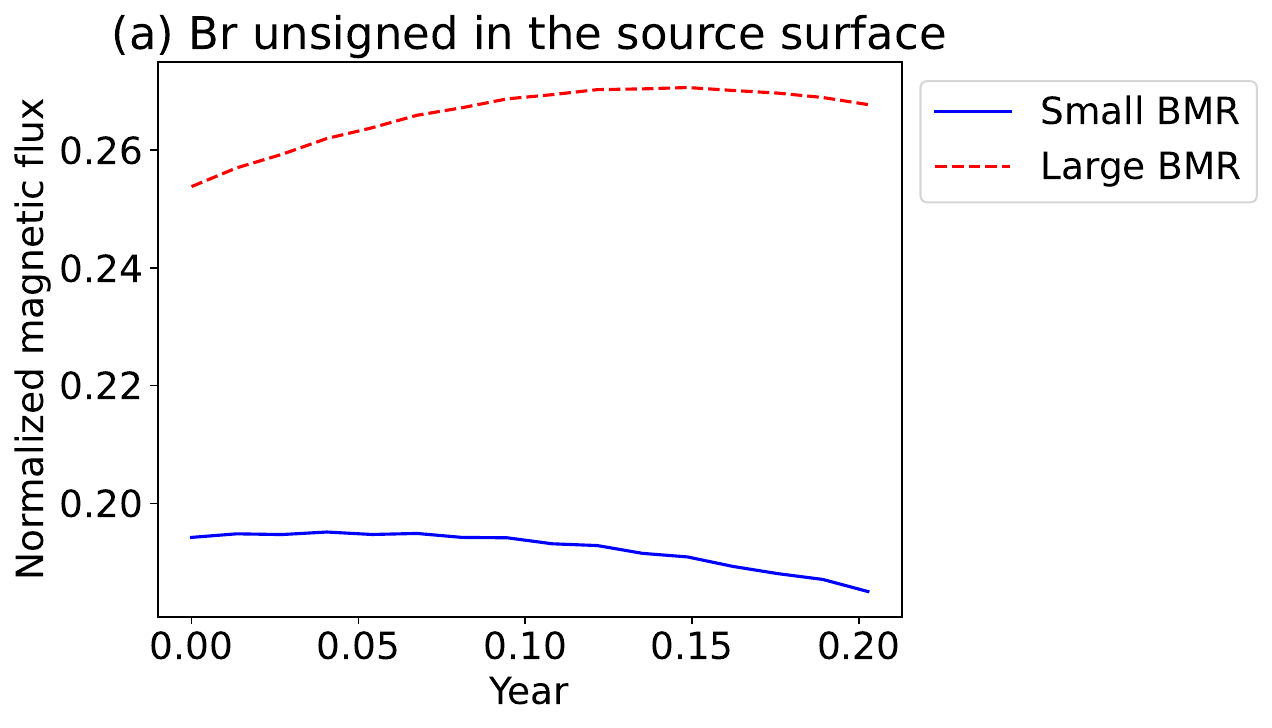}
   \includegraphics[width=.45\columnwidth]{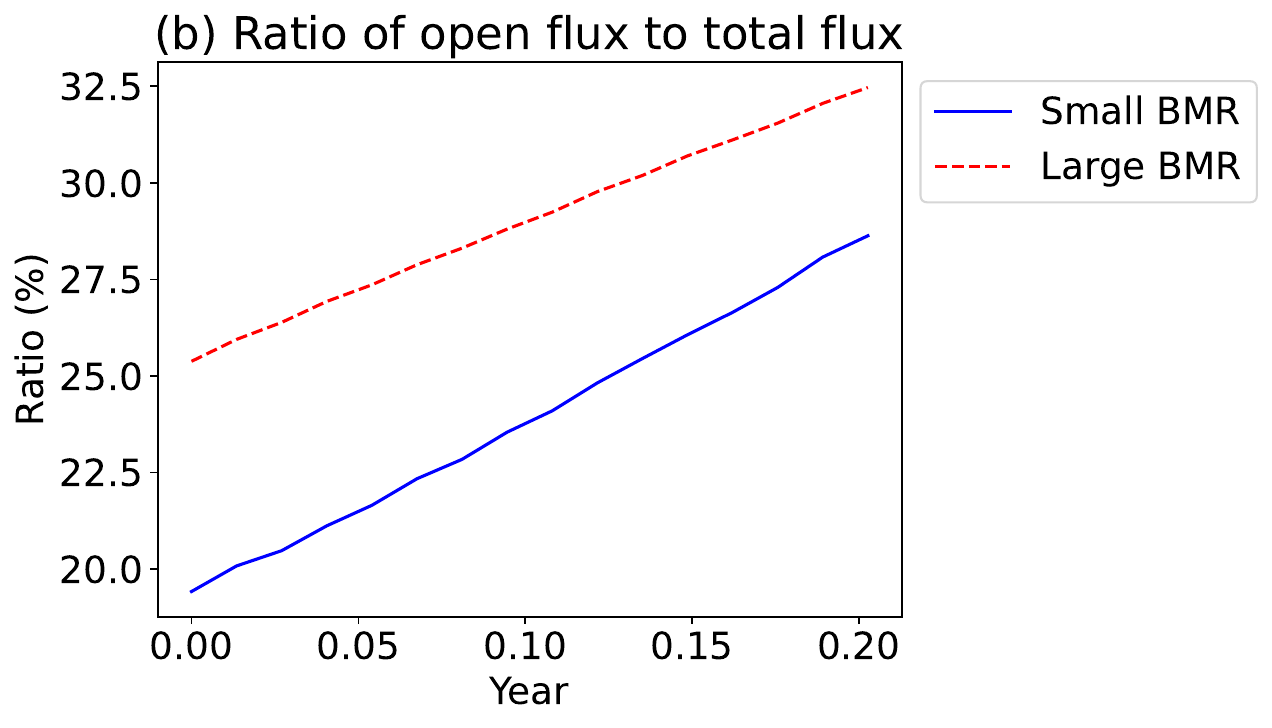}
  \caption{
  \rev{Time evolution of the amount of the magnetic flux in the source surface (a) and the ratio of the open flux at the source surface (b) when a small BMR (solid blue lines) or a large BMR (dashed red lines) is placed. All values are normalized by the initial value of the total unsigned magnetic flux in the photosphere.}
  } 
  \label{Fig:rad_plot}
 \end{center}
\end{figure}

\rev{We investigated how the temporal evolution of open flux depends on the magnetic flux intensity of BMRs when they emerge into a pre-existing open field structure that has already expanded to middle latitudes. The new BMRs were placed at the same latitudes and longitudes as the initial BMR used to generate the background. This configuration is assumed based on the conditions observed during CR 2155 and CR 2156.
Figure \ref{Fig:rad_plot} shows the time evolution of the normalized total open flux (a) and the ratio of the open flux to the total flux (b). At the initial condition ($t=0$), the open flux ratio is 25.3\% for the large BMR case (red lines), whereas it is 19.4\% for the small BMR case (blue lines). This indicates that the large BMR makes the magnetic field open more effectively by a factor of approximately 1.3 compared to the small BMR when the background open field has already expanded to middle latitudes (around $30^{\circ}$). As shown in Figure \ref{Fig:rad_plot}(a), the total open flux increases over time only in the case of the large BMR. This difference occurs because the amount of photospheric magnetic flux cancellation depends on the size of the BMR. Conversely, Figure \ref{Fig:rad_plot} (b) shows that the open flux ratio increases with a similar trend in both cases.}

\begin{figure}[ht]
 \begin{center}
   \includegraphics[width=.8\columnwidth]{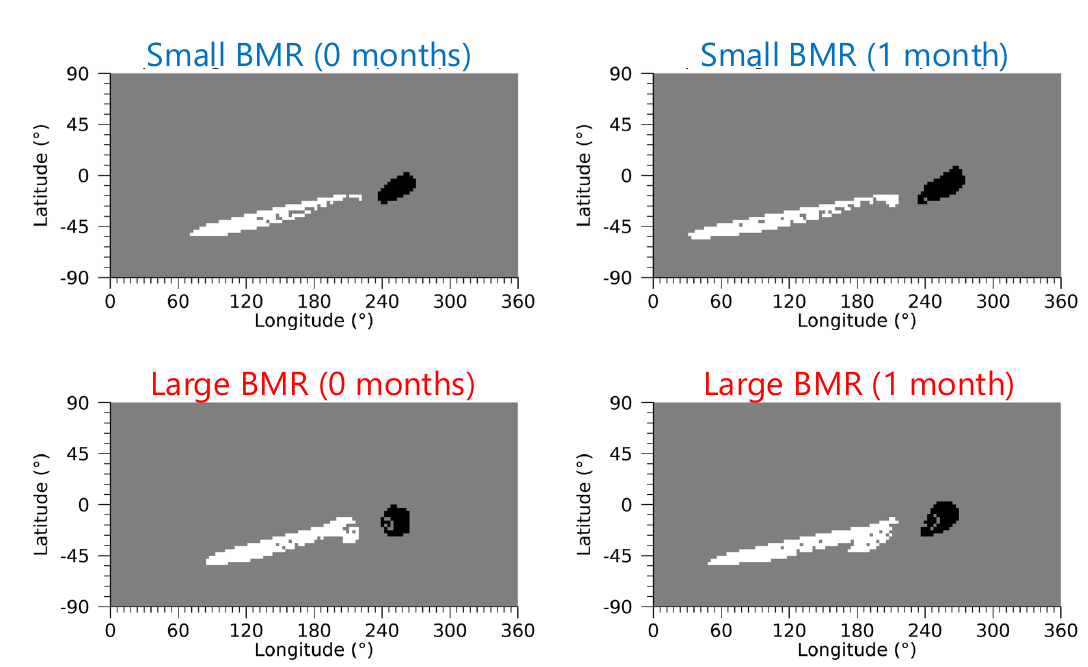}
  \caption{
   \rev{The open flux footpoints for configuration \#1. The upper panel shows the open flux footpoints when a small BMR is placed, and the lower panel shows that when a large BMR is placed. The black and white regions represent negative and positive open area.}
  } 
  \label{Fig:rad_foot}
 \end{center}
\end{figure}

\rev{Figure \ref{Fig:rad_foot} shows the open flux footpoint maps for the small BMR (upper panels) and the large BMR (lower panels). The left and right panels correspond to the initial state and the state after one month, respectively. We focus on the positive open field (white) associated with the following sunspot polarity. The magnetograms show that the pre-existing open field merges with the open field originating from the new BMR. Comparing the two cases, the positive open field region in the large BMR case is narrower in the longitudinal direction but more extended in the latitudinal direction than in the small BMR case. This trend persists even after one month. Furthermore, the positive open field remains clearly visible around the BMR region in the large BMR case.}

\subsubsection{Dependence on the configurations of BMRs with Initial Background Magnetic Fields (Configuration $\#$2)}\label{sec:ch_diff}
\begin{figure}[ht]
 \begin{center}
   \includegraphics[width=.45\columnwidth]{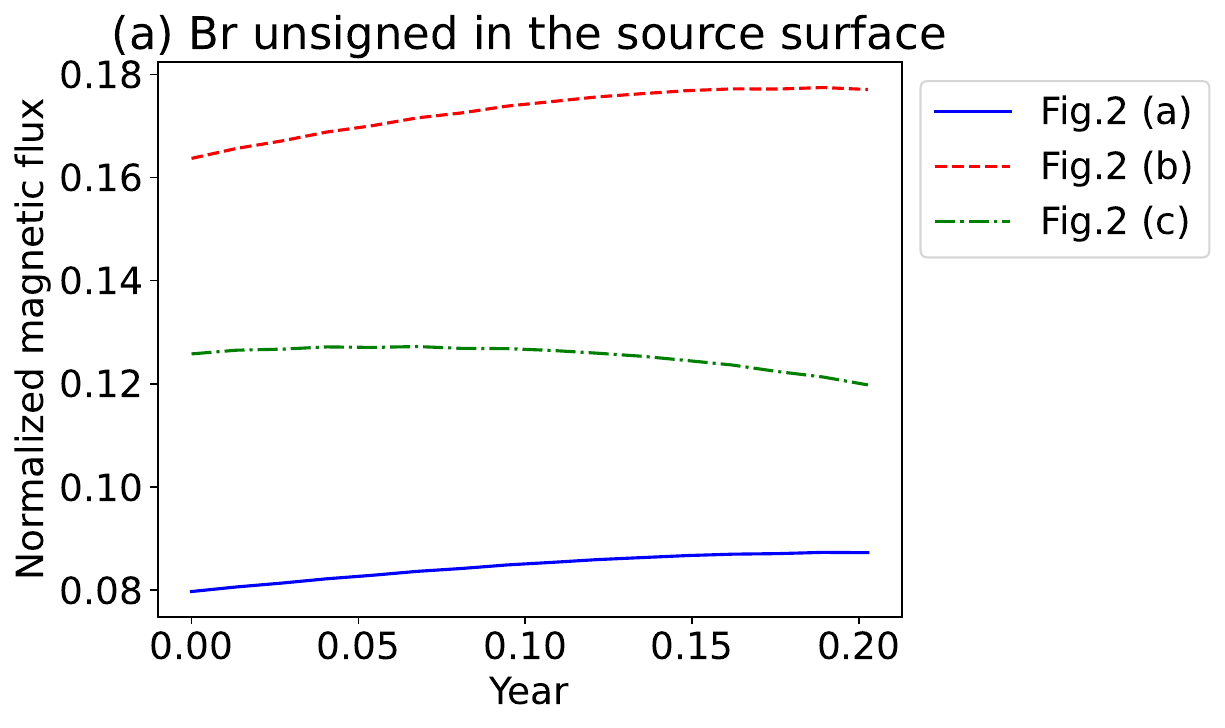}
   \includegraphics[width=.45\columnwidth]{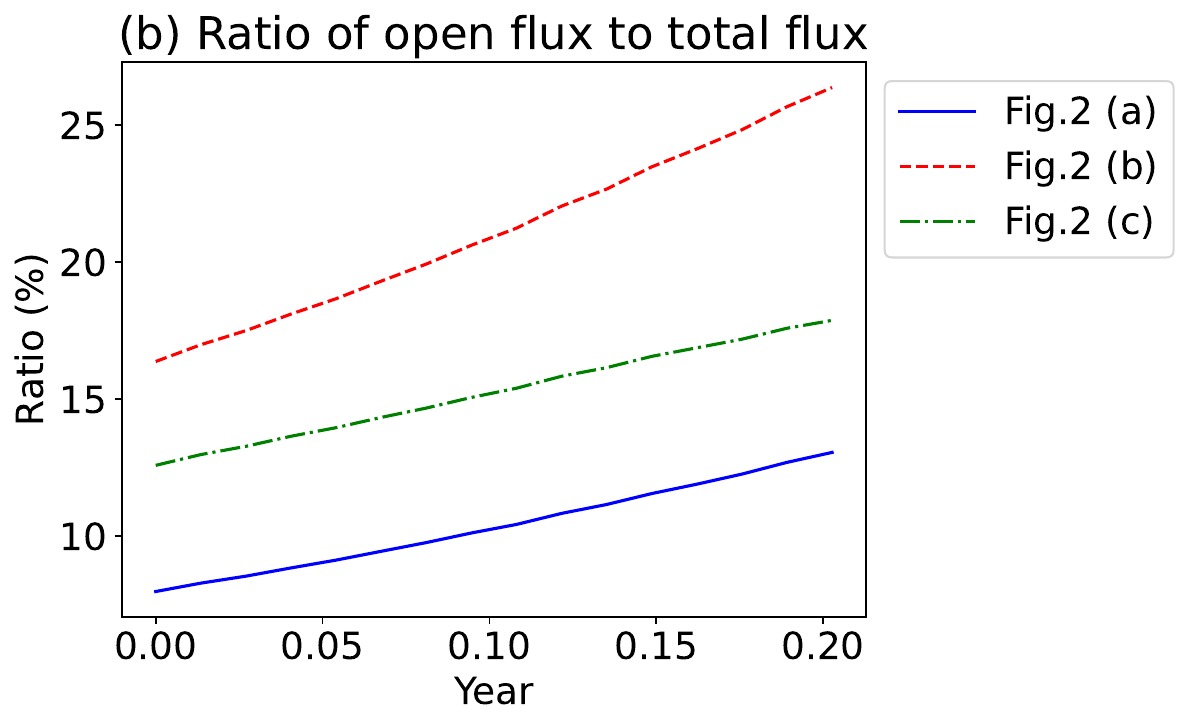}
  \caption{
  \rev{Time evolution of the amount of the magnetic flux in the source surface (a) and the ratio of the open flux at the source surface (b). The blue, red, and green lines represent the result for Configuration \#2 (a), (b), and (c) in Figure \ref{fig:sun_tilt}, respectively. All values are normalized by the initial value of the total unsigned magnetic flux in the photosphere.}
  } 
  \label{Fig:diff_plot}
 \end{center}
\end{figure}

\rev{Next, we investigate the case of configuration \#2 (a)-(c) as shown in Figure \ref{fig:sun_tilt}. Figure \ref{Fig:diff_plot} shows the time evolution of the normalized total open flux (a) and the ratio of the open flux to the total flux (b). At the initial condition ($t=0$), the open flux ratios are 7.97\%, 16.4\%, and 12.6\% for the Configuration (a; blue lines), (b; red lines), and (c; green lines) in Figure \ref{fig:sun_tilt}, respectively. After one month, the ratio of open flux increased by a factor of 1.22 for configurations (a) and (b), and 1.17 for Configuration (c). The total open flux in panel (a) decreases after one month in the case of Configuration (c). These trends are consistent with the results without initial background magnetic fields. }

\begin{figure}[ht]
 \begin{center}
   \includegraphics[width=.98\columnwidth]{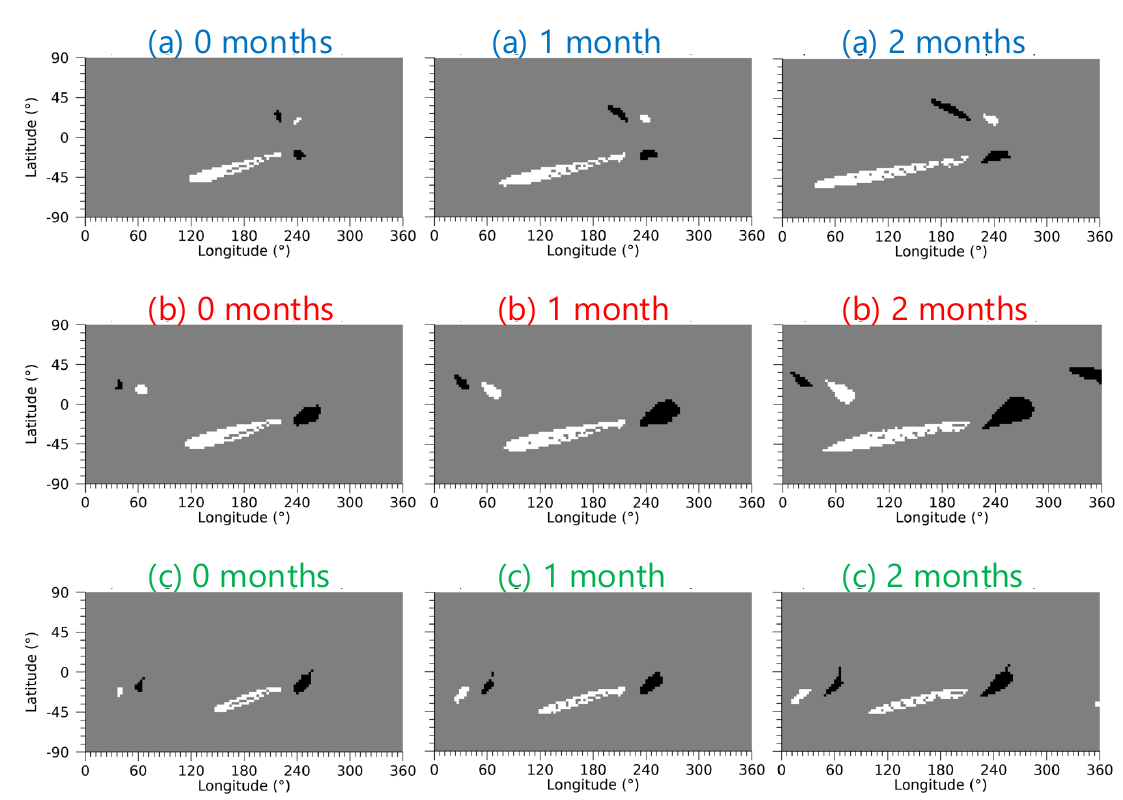}
  \caption{
   \rev{The open flux footpoints for configuration \#2. The upper panels show the open flux footpoints for Configuration \#2 (a), the middle panels show that for Configuration \#2 (b), and the lower panels show that for Configuration \#2 (c). The black and white regions represent negative and positive open area. \red{Note that open field regions with negligible magnetic field strength are excluded.}}
  } 
  \label{Fig:diff_foot}
 \end{center}
\end{figure}

\rev{Figure \ref{Fig:diff_foot} shows the open flux footpoint maps for configurations (a), (b), and (c) from top to bottom. The panels from left to right correspond to the initial state, the state after one month, and the state after two months, respectively.
First, we focus on the overall distribution of the positive open field. Configuration (a) shows the most significant expansion in the longitudinal direction, while Configuration (b) shows the expansion in the latitudinal direction. In contrast, in Configuration (c), the positive open field in the southern hemisphere shows less expansion compared to the other two cases.
The expansion characteristics of the negative open field (black) also differ depending on the configuration. In Configuration (a), the open field originating from the following polarity of the BMR placed in the northern hemisphere expands longitudinally. In Configuration (b), the open field originating from the leading polarity of the BMR placed at $230^{\circ}$ longitude expands the most, reaching near $0^{\circ}$ latitude. Configuration (c) shows that the negative open field remains in a more concentrated region compared to the other two configurations, similar to the positive open field.}

\FloatBarrier
\subsection{Discussion on the simulation results}\label{sec:discussion_simu}

\rev{Before this study, we first confirmed that the time evolution of open flux is governed by the fundamental properties of BMRs, consistent with pioneering works \citep{2002JGRA..107.1302W, 2002SoPh..207..291M}. 
Our preliminary simulations verified that larger tilt angles enhance the open flux, while lower latitudes help maintain it. 
Furthermore, we confirmed that the increase in open flux is driven by the equatorial dipole component, which is amplified effectively by differential rotation rather than meridional flow.
Given that these fundamental mechanisms are consistent with established theory, we focus our main discussion on the interaction between emerging BMRs and the pre-existing magnetic fields (Section \ref{sec:result2}).
Specifically, we discuss how the magnetic intensity (Section \ref{sec:rad_dis}) and geometric configuration (Section \ref{sec:ch_dis}) of sunspots affect the evolution of open flux in a realistic solar environment.}

\subsubsection{Effect of BMR Strength and Initial Background Magnetic Fields on Open Flux}\label{sec:rad_dis}
\rev{Regarding the dependence on BMR strength (Section \ref{sec:ch_rad}), we found that the large BMR \red{generates} the approximately 1.3 times \red{more total open flux} than the small BMR. This result depends on the dipole magnetic moment of the BMR. While the equatorial dipole component remains almost unchanged between the two BMRs, the axial dipole component of the small BMR is half that of the large BMR. Consequently, the total dipole moment in the large BMR case is 1.26 times larger than that in the small BMR case. Furthermore, since a small BMR is canceled more easily by the pre-existing open field (\citealp{1994SoPh..150....1S}), their contribution to the axial dipole moment is limited \citep{2002JGRA..107.1302W}. This likely explains why significant open fields remained in the surroundings of the BMR only in a large BMR case. Therefore, this result suggests that the emergence of a BMR with strong magnetic flux near an open field region leads to an increase in open magnetic flux \red{when the BMR has a polarity that allows it to merge with the pre-existing open field}.}

\subsubsection{Effect of BMR Configuration and Initial Background Magnetic Fields on Open Flux}\label{sec:ch_dis}

\rev{In the presence of an initial background magnetic field, the evolution of the open field is governed by the competition between the formation of closed loops and the merging with the existing open field (Section \ref{sec:ch_diff}). The latter process is driven by interchange reconnection, where the footpoint of an open field line is exchanged with a neighboring closed loop, allowing the coronal hole to move or expand effectively \citep{2004ApJ...612.1196W}.
In Configuration (a) of Figure \ref{Fig:diff_foot}, the open field expanded in the longitudinal direction. The emergence of BMRs at the same longitude facilitates the formation of closed loops in the latitudinal direction \citep{2002JGRA..107.1302W}. These closed loops prevent the open field from extending equatorward. However, the pre-existing background field merges with the same-polarity component of the BMRs via diffusion. It is considered that the open flux extends primarily in the longitudinal direction due to the latitudinal closed loops.
In Configuration (b) of Figure \ref{Fig:diff_foot}, the expansion of the positive open fields is the most significant, particularly in the latitudinal direction. Because the BMRs are widely separated in both longitude and latitude, this configuration inhibits the formation of closed loops between the northern and southern BMRs. Therefore, the polarity of the placed BMRs leads to interchange reconnection with the initial open fields, facilitating the footpoint exchanges described by \citet{2004ApJ...612.1196W}. As a result, the positive open field extends from the middle latitudes toward the equator.
In Configuration (c) of Figure \ref{Fig:diff_foot}, the open field was not significantly extended compared to the other two cases. We attribute this to the placement of BMRs at the same latitude creates the longitudinal closed loops, which inhibit the expansion of the open field area in the latitudinal direction. The presence of an initial background magnetic field between the two BMRs made it easier for the magnetic field to close, unlike in Configuration \red{(b)}.}

\red{These discussions are consistent with the quantitative argument that maximizing the open flux requires maximizing the global dipole strength. The configurations that facilitate closed loop formation (Configurations (a) and (c)) reduce the net dipole moment through magnetic cancellation, whereas Configuration (b), which minimizes magnetic cancellation, preserves a larger dipole moment. Thus, the temporal variations in the open flux can be fundamentally attributed to the modulation of the global dipole strength by the configurations of BMRs.}

\FloatBarrier
\section{Applications to the observation in 2014}\label{sec:result_obs}
\subsection{Result: Rapid IMF Increase Event Observed in 2014}\label{sec:result:IMFobs} 

\begin{figure}[ht]
 \begin{center}
  \includegraphics[width=.7\columnwidth,
  trim=4mm 6mm 4mm 4mm,
  clip]{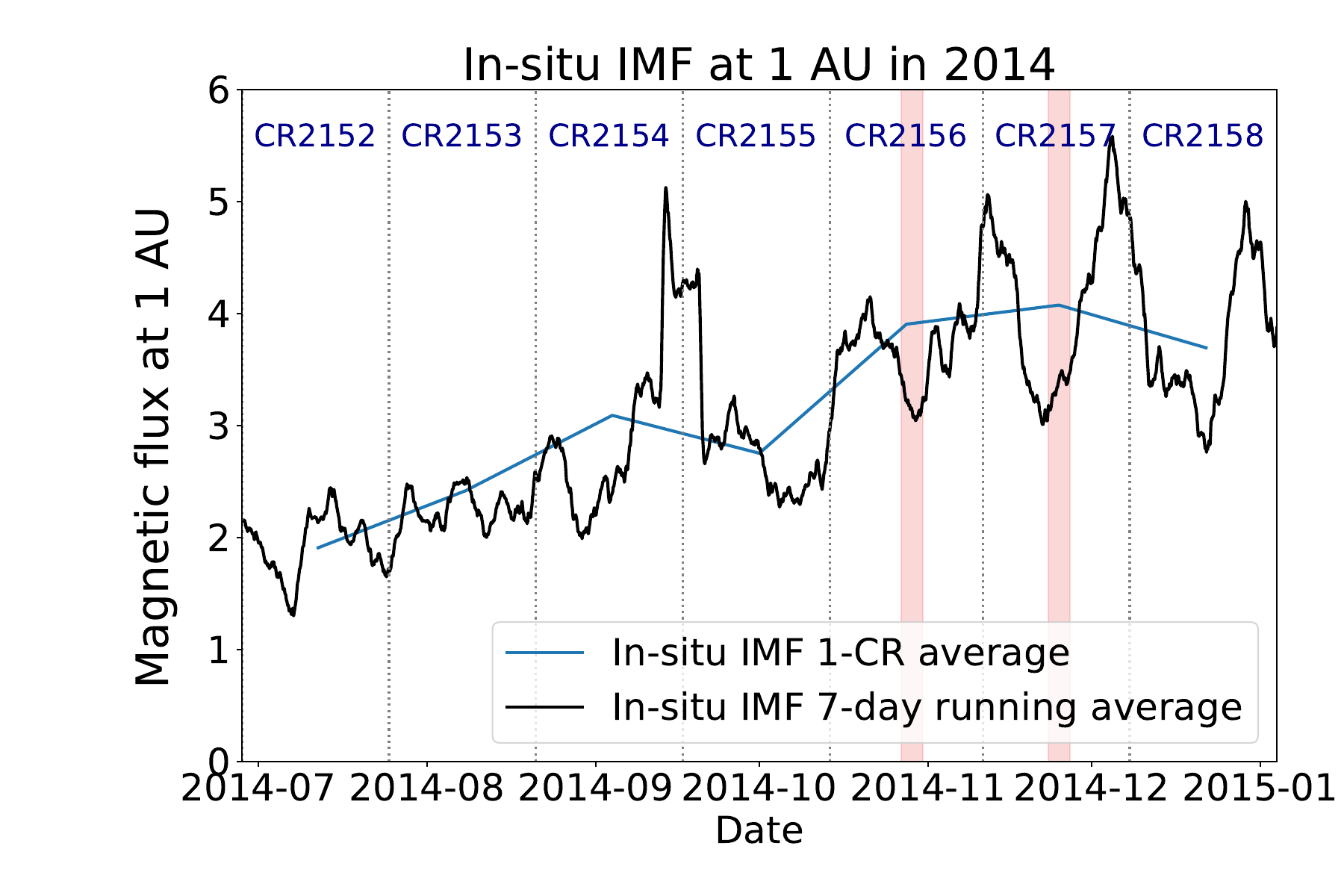}\\
  \includegraphics[width=.7\columnwidth,
  trim=4mm 6mm 4mm 4mm,
  clip]{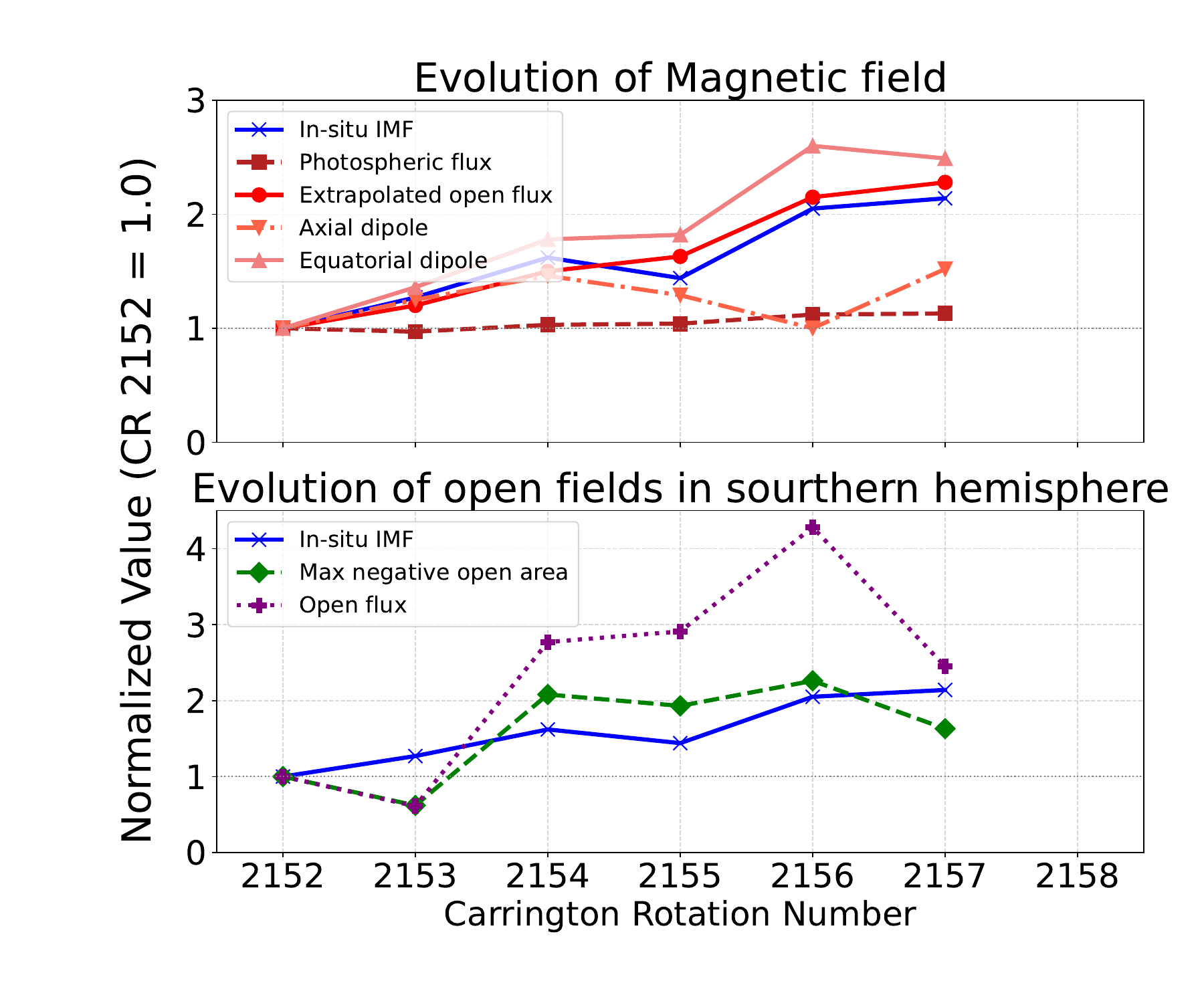}
  \caption{
  The upper panel shows the time evolution of in-situ IMF in 2014. The blue line shows a 1-month average, and the black line shows a 7-day running average. The red shading in CR 2156 indicates the period 5 days after AR12192 passed through the meridional line on the photosphere. The red shading in CR 2157 indicates the period 5 days after the same active region \red{reappeared} on the photosphere in the next Carrington rotation at the passage to the disk center. \rev{The lower two panels show the time variations of parameters summarized in Table \ref{tab:obs}. The horizontal axis is the Carrington rotation number, and the vertical axis is the normalized value by the value in CR 2152.}
  } 
  \label{Fig:wavelet}
 \end{center}
\end{figure}

Figure \ref{Fig:wavelet} shows the temporal evolution of the in-situ IMF in 2014 and solar magnetic field components. In the upper figure, The red shading areas show 3-6 days after a huge active region numbered 12192 (AR 12192) passed through the central meridian. This time delay corresponds to the travel time for a change in open magnetic flux generated at the solar surface to reach Earth via the solar wind. \rev{The in-situ IMF increased by a factor of 2.14 from CR 2152 to CR 2157. Based on Figure \ref{Fig:wavelet}, this increasing trend can be divided into three phases: (i) a gradual increase by a factor of approximately 1.27 per rotation from CR 2152 to CR 2154, (ii) a rapid increase by a factor of 1.42 from CR 2155 to CR 2156, and (iii) a slight increase by a factor of 1.05 from CR 2156 to CR 2157.} Note that the increase in IMF seen at the end of CR 2154 appears to be related to series of coronal mass ejections (CMEs) directed to the Earth that occurred on 10-12 September 2014 (\citealp{2017SoPh..292..142W}; \citealp{2023A&A...675A.136M}; \citealp{2024ApJ...965..151H}).

\rev{The lower two panels in Figure \ref{Fig:wavelet} and Table \ref{tab:obs} show the time variation of the in-situ IMF, photospheric magnetic fields, extrapolated open flux, solar axial and equatorial dipole flux, and maximum negative open area and flux from CR 2152 to CR 2157. Figure \ref{Fig:wavelet} shows the normalized value by the value in CR 2152.} 
\rev{The total photospheric magnetic field (B) remained relatively stable during this period, with a maximum-to-minimum ratio of only 1.17. This stands in contrast to the in-situ IMF (A) and the open flux ratio (D), both of which increased by a factor of more than 2. The equatorial dipole flux (F) and the extrapolated open flux (C) have strong correlations with the variations in the in-situ IMF, with correlation coefficients of 0.98 and 0.97, respectively. In contrast, the axial dipole flux (E) shows a weak correlation of 0.37 with the in-situ IMF. However, notably from CR 2156 to CR 2157, the equatorial dipole flux decreased while the axial dipole flux increased. This suggests that the growth of the axial dipole component contributed to the continued increase in the IMF during this specific interval.} 

\rev{The bottom panel in Figure \ref{Fig:wavelet} and the bottom two rows (G, H) in Table \ref{tab:obs} shows the maximum negative open area and the corresponding open magnetic flux from CR 2152 to CR 2157 to examine the coronal hole (open field) expanding in the southern hemisphere. 
The negative open area expanded by a factor of 2.0 from CR 2152 to CR 2154, and the total flux of the largest open region increased by a factor of 2.8 from CR 2152 to CR 2154 and by a factor of 1.5 from CR 2155 to CR 2156.
First, the area expansion observed in CR 2153 is attributed to the formation of the coronal hole in the southern hemisphere.
Furthermore, in CR 2154 and CR 2155, sunspot groups were observed at longitudes similar to that of the giant active region (AR 12192) observed in CR 2156.
Consequently, the observed increase in open flux is likely due to the incorporation of the strong magnetic fields surrounding these active regions into the open field structure.}

\begin{figure}[ht]
 \begin{center}
   \includegraphics[width=.8\columnwidth]{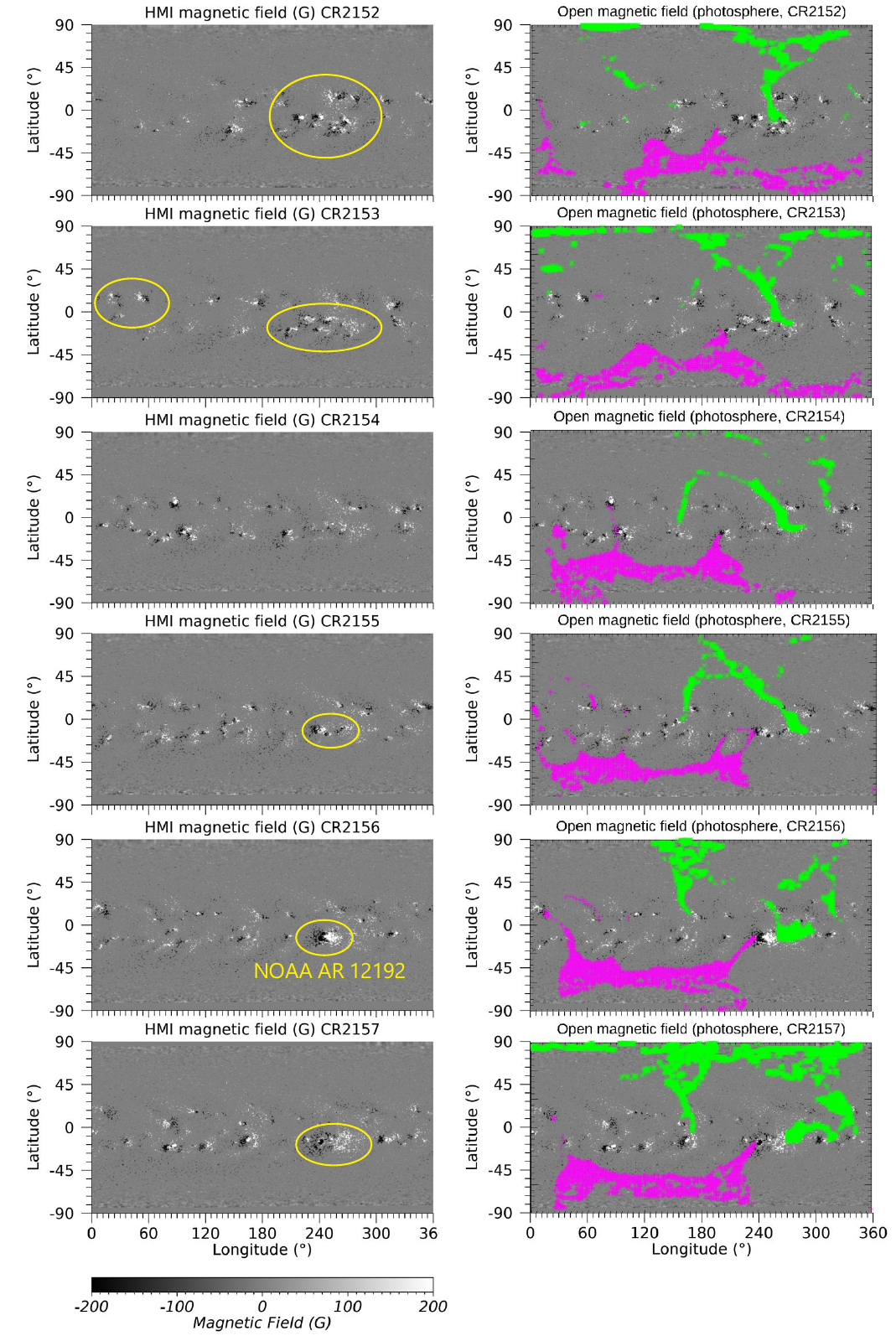}
  \\ 
  \caption{
  The left panels show synoptic maps \rev{from CR 2152 to CR 2157}, and the right panels show the open flux footpoint, respectively. The yellow circles indicate characteristic sunspot groups discussed in the text. The green represents outward open flux, while the pink is inward flux.
  } 
  \label{Fig:obs}
 \end{center}
\end{figure}

Figure \ref{Fig:obs} show synoptic maps of magnetic fields and footpoints \rev{from CR 2152 to CR 2157}, respectively. \rev{Characteristic sunspot groups discussed in the text are marked with yellow circles.} In CR 2152, a number of sunspot groups of similar size with similar total magnetic flux are spatially distributed in certain ranges of longitudes (around $150^{\circ}$ and $260^{\circ}$) at similar latitudes ($10^{\circ}-20^{\circ}$). \rev{This configuration is similar to case (a) shown in Figure \ref{fig:sun_tilt}. In CR 2153, the sunspot groups are spatially separated as marked, resembling Configuration \red{\#2} (b). In CR 2155, a sunspot group appears at longitude $240^{\circ}$ (approximately $4.6\times10^{22}$ Mx), corresponding to the region where AR 12192 emerges in CR 2156.} In CR 2156, AR 12192 (approximately $1.3\times10^{23}$ Mx) newly appeared as marked by the yellow circle in Figure \ref{Fig:obs}. The sunspot group is tilted by $12.7^{\circ}$ and the size is more than twice that of the other sunspot groups in this synoptic map. In CR 2157, the magnetic field of the huge sunspot group diffuses as marked. For the open flux footpoints, In CR 2156 and CR 2157, much of the open flux extends from the areas around AR 12192, and less from the center of AR 12192.
During the period of red shading areas in the bottom panel of the Figure \ref{Fig:wavelet}, which is 5 days after AR 12192 located at the disk center, the IMF weakens in each Carrington rotation. The IMF before and after the shaded periods is approximately 1.5 times larger than within them. Based on these results, and assuming a magnetic field propagation time of 5 days from the Sun to 1 AU, we find that periods with less open field extending from the photosphere correspond to weaker in-situ IMF at 1 AU. Conversely, the in-situ IMF is stronger when there is significant open flux around the active region extending to 1 AU.
\rev{Figure \ref{Fig:obs} also shows a significantly expanded negative coronal hole (pink) from CR 2154 to CR 2157, in consistent with (G) in Table \ref{tab:obs}.}

\begin{table}[!ht]
    \centering
    \caption{Values and variations of the in-situ IMF, photospheric magnetic field, the open flux extrapolated from the PFSS model, solar axial and equatorial dipole strength\rev{, the maximum negative open field area, and the open flux in the maximum open region from CR 2152 to CR 2157. The negative open field represents the negative polarity region extending from the coronal hole in the sourthern hemisphere.}} 
    \begin{tabular}{lcccccc}
    \hline
        Carrington Rotation Number & CR 2152 & CR 2153 & CR 2154 & CR 2155 & CR 2156 & CR 2157 \\ \hline \hline
        (A) In-situ IMF ($10^{22}$ Mx) & 5.37 & 6.80 & 8.69 & 7.74 & 11.0 & 11.5 \\ \hline
        (B) Photospheric magnetic field ($10^{23}$ Mx) & 6.99 & 6.78 & 7.19 & 7.30 & 7.85 & 7.92 \\ \addlinespace
        (C) Extrapolated open flux ($10^{22}$ Mx) & 2.02 & 2.43 & 3.04 & 3.30 & 4.35 & 4.61 \\ \addlinespace
        (D) Ratio of open flux (C)/(D) (\%) & (2.89) & (3.58) & (4.23) & (4.51) & (5.54) & (5.82) \\ \addlinespace
        (E) Axial dipole flux ($10^{22}$ Mx)  & 0.65 & 0.81 & 0.96 & 0.84 & 0.65 & 0.99 \\ \addlinespace
        (F) Equatorial dipole flux ($10^{22}$ Mx)  & 1.01 & 1.37 & 1.80 & 1.84 & 2.63 & 2.51 \\ \hline
        (G) Max negative open area ($10^{21} \mathrm{cm^{2}}$) & 0.73 & 0.45 & 1.52 & 1.41 & 1.65 & 1.19 \\ \addlinespace
        (H) Open flux in above region ($10^{22}$ Mx) & 0.65 & 0.40 & 1.81 & 1.90 & 2.80 & 1.60 \\ \hline
    \end{tabular}
    \label{tab:obs}
\end{table}

\FloatBarrier
\subsection{Discussion: The cause of the IMF significant increase in Cycle 24}\label{sec:discussion:IMFobs}
From the results shown in Section \ref{sec:result:IMFobs}, the average IMF intensity increased by a factor of 2.14 from CR 2152 to CR 2157, and the ratio of open flux increased by a factor of 2.0 during this period.
\rev{\citet{2024ApJ...965..151H} suggested a potential link between the IMF rise and the emergence of this largest active region, as well as the southern coronal hole. However, they did not clarify the quantitative contribution or the physical mechanism of how these features increased the open flux. We discuss the causes of the IMF rapid increase by combining our simulation results with the detailed observational analysis. The increase in the open flux ratio can be attributed to three primary drivers related to Section \ref{sec:result2}: (i) characteristics of BMRs (e.g., tilt angle and magnetic field strength) and their diffusion, (ii) the relative configuration of emerging sunspot groups, and (iii) the interaction between new BMRs and the pre-existing magnetic fields. Below we discuss how these factors contributed to the three phases of the IMF increase observed in 2014.
}
\rev{From Section \ref{sec:result:IMFobs}, the IMF rapid increase can be divided into three phases: (i) a steady increase from CR 2152 to CR 2154, (ii) a rapid surge from CR 2155 to CR 2156, and (iii) a sustained high level from CR 2156 to CR 2157.
The initial increase from CR 2152 to CR 2154 can be explained by the diffusion and configuration of BMRs.
In CR 2152, sunspot groups are placed at similar longitudes. This spatial configuration is close to Configuration \red{\#2} (a). The simulation showed that this configuration increases the open flux ratio by a factor of 1.22 after one month (Section \ref{sec:ch_diff}). This value is consistent with the observed IMF increase rate of approximately 1.27 per rotation during this phase.}

\rev{In CR 2153, the sunspot distribution shifted to a Configuration \red{\#2} (b). Although our simulation suggests that Configuration \red{\#2} (b) has the potential to double the open flux ratio compared to Configuration \red{\#2} (a), the observed increase in CR 2153 was limited.
This discrepancy can be explained by the time required for footpoint exchange. The configuration of sunspot groups in CR 2153 leads to merging the BMR flux with the background coronal hole, extending it latitudinally.
While the maximum negative open area in Table \ref{tab:obs} (G) shows a temporary fluctuation, the footpoint maps (Figure \ref{Fig:obs}) indicate that the southern open field region began to expand globally in CR 2153.
It is likely that the open flux was maintained during the rapid expansion of the coronal hole in the southern hemisphere (footpoint exchange mechanism described by \citealp{2004ApJ...612.1196W}).
This implies that the configuration of BMRs is favorable for increasing in the total open flux, but the increase in open flux of coronal holes requires time for magnetic restructuring.
Furthermore, frequent SEP events (see Section \ref{sec:result_obs}) suggest that associated CME activity also contributed to expand the open fields (a large coronal hole) in sourthern hemisphere.
Consequently, in CR 2154, the southern hemisphere open fields (forming a large coronal hole) expanded significantly in terms of both area and magnetic flux.}

\rev{The rapid increase in IMF from CR 2155 to CR 2156 is attributed to the characteristics of BMRs.
Comparing CR 2155 and CR 2156, a BMR appeared at the same location (longitude $\sim 240^{\circ}$), but the total magnetic flux of the active region in CR 2156 (AR 12192) was approximately 2.8 times larger than that in CR 2155.
Our simulation on BMR strength (Section \ref{sec:ch_rad}) showed that a Large BMR (with 2 times the magnetic flux of a Small BMR) has a 1.3 times larger ratio of open flux.
Applying this result to the observation, it is considered that the 2.8 times difference in magnetic flux intensity can explain the observed 1.42 times increase in the IMF. 
Furthermore, we focus on the footpoint configuration in CR 2156. As shown in the open flux footpoint map in Figure \ref{Fig:obs}, a positive open field is observed near the leading sunspot of AR 12192. This distribution is close to the footpoint pattern shown in Figure \ref{Fig:diff_foot}. This suggests that the giant sunspot group formed a global magnetic structure that can be assumed to be a configuration where the open field of the BMR merges with the adjacent coronal hole (configuration \red{\#1}).}

\rev{The IMF increased by a factor of 1.05 from CR 2156 to CR 2157. This phase corresponds to the diffusion of the giant active region AR 12192. Our results in Section \ref{sec:ch_rad} indicate that the diffusion of magnetic flux (Factor iii) plays a critical role in this process. A single BMR with a large tilt angle can increase the open flux ratio by approximately 5\% over one month. This value is consistent with increase in in-situ IMF. As shown in the simulation of Configuration \red{\#1} over two months (Section \ref{sec:ch_rad}), the open field formed around a large BMR merges with initial open fields and it expands. This process contributes to the reinforcement of the solar axial dipole flux. Table \ref{tab:obs} (E) shows an increase in the axial dipole flux from CR 2156 to CR 2157. Thus, it is considered that the diffusion of the giant BMR leads to an increase in IMF in this period.}

\section{Conclusion}\label{sec:conclusion}
In this study, we investigated the generation and formation of the open flux during solar maximum when the open flux footpoints are complex, and simulated the open flux from around the sunspot groups. 
\rev{We elucidated the mechanism of the rapid IMF increase observed in 2014 by applying theories derived from the SFT model with BMRs to the observed magnetic structures.
We identified three key factors of BMRs essential for explaining the ratio of open flux increase: (i) the characteristics of BMRs (intensity and tilt angle), (ii) the relative configuration between sunspot groups, and (iii) the interaction with the pre-existing open fields. In the presence of an initial background magnetic field, the relative configuration of BMRs can either inhibit open flux expansion by forming closed loops or promote it through favorable connections.}

\rev{Based on our simulation results, the rapid IMF increase observed from CR 2152 to CR 2157 can be explained  with three factors for increasing the ratio of open flux and devided three phases:
\begin{enumerate}
    \item CR 2152-2154: The gradual IMF increase is driven by the sunspot configuration and the expansion of the southern coronal hole. While the Configuration \#2 (a) (CR 2152) gradually increases the open flux over time, the Configuration \#2 (b) (CR 2153) facilitates latitudinal expansion of negative open fields. Our analysis suggests that the delay in open flux increase in CR 2153 is due to the time required for footpoint exchange with the sourthern open fields, which resulted in a significant expansion of coronal holes in CR 2154.
    \item CR 2155-2156: The rapid IMF increase is primarily caused by the magnetic intensity of the BMR. We showed that the giant sunspot group (AR 12192), which had approximately 2.8 times the magnetic flux of typical BMRs, significantly enhanced the open flux ratio. It is considered that this results can explain the observed 1.42 times increase in the IMF.
    \item CR 2156-2157: The slight IMF increase corresponds to the diffusion phase of the giant sunspot group (AR 12192). This increase can be explained by the diffusion of the BMR with a high tilt angle, and merging and expanding the southern open fields contribute to the reinforcement of the axial dipole field.
\end{enumerate}
}

\rev{These results emphasize, both observationally and theoretically, that the temporal variations of open flux and IMF during solar maximum cannot be explained by BMR characteristics alone. It is necessary to consider the interaction between the sunspot groups and the surrounding large coronal holes (open fields).
}

\section*{Acknowledgements}
The data courtesy of NASA/SDO and the HMI science team. We thank the Joint Science Operations Center at Stanford for providing HMI synoptic maps. We acknowledge the use of NASA/GSFC's Space Physics Data Facility's OMNIWeb service and OMNI data\red{, and the Solar Influences Data Analysis Center (SILSO), Royal Observatory of Belgium, for providing the sunspot number series}. We thank the anonymous referee for the important and valuable comments. This work was supported by JST SPRING, Grant Number JPMJSP2108, JSPS KAKENHI Grant No. JP20KK0072, JP21H01124, JP21H04492, and JP25KJ1031, and the grant of OML Project by the National Institutes of Natural Sciences (NINS program No. OML032402).
    




\bibliography{sample701}{}
\bibliographystyle{aasjournalv7}



\end{document}